\documentclass[journal,10pt]{IEEEtran}
\usepackage{etex}
\pdfoutput=1
\usepackage{tabularx} 
\usepackage{pdfpages}
\usepackage{algorithm,algpseudocode,graphicx}
\usepackage{tikz}
\usetikzlibrary{shapes,arrows}
\usepackage{pstricks}
\usepackage{pst-node,pst-blur}
\usetikzlibrary{arrows}
\usepackage{amsfonts}
\usepackage{tabularx}
\usepackage{enumerate}
\usepackage{subfigure}
\usepackage{amssymb}
\usepackage{booktabs} 
\usepackage{multirow} 
\usepackage{epsfig}
\usepackage{epstopdf}
\usepackage{array}

\usepackage[noadjust]{cite}

\newtheorem{remark}{Remark}

\newtheorem{proposition}{Proposition}

\usepackage[cmex10]{amsmath}
\usepackage[font=small,skip=1.5pt]{caption}
\allowdisplaybreaks
\begin{document}

\title{Utility Regions for DF Relay in OFDMA-based Secure Communication with Untrusted Users\thanks{Manuscript received June 24, 2017; accepted July 16, 2017. This work has been supported by the Department of Science and Technology under Grant no. SB/S3/EECE/0248/2014. The associate editor coordinating the review of this paper and approving it for publication was K. Tourki.}}
\author{Ravikant Saini, {\em{Member, IEEE}}, Deepak Mishra, {\em{Student Member, IEEE}}, and Swades De, {\em{Senior Member, IEEE}}
\thanks{R. Saini is with the Department of Electrical Engineering, Shiv Nadar University,  Uttar Pradesh 201314, India (email: ravikant.saini@snu.edu.in).}
\thanks{D. Mishra and S. De are with the Department of Electrical Engineering and Bharti School of Telecommunication,  Indian Institute of Technology Delhi, New Delhi 110016, India (e-mail: \{deepak.mishra, swadesd\}@ee.iitd.ac.in).}
\thanks{Digital Object Identifier xxxxxxxxxxxxxxxxx}
}

\maketitle

\begin{abstract}
This paper investigates the utility of a trusted decode-and-forward relay in OFDMA-based secure communication system with untrusted users. For deciding whether to use the relay or not, we first present optimal subcarrier allocation policies for direct communication (DC) and relayed communication (RC). Next we identify exclusive RC mode, exclusive DC mode, and mixed (RDC) mode subcarriers which can support both the modes. For RDC mode we present optimal mode selection policy and a suboptimal strategy  independent of power allocation which is asymptotically optimal at both low and high SNRs. Finally, via numerical results we present insights on relay utility regions.
\end{abstract}

\IEEEpeerreviewmaketitle
\bstctlcite{IEEEexample:BSTcontrol} 
\section{Introduction}
\label{sec_introduction}
With growing number of users, utilization of friendly relays for providing secure communication to cell-edge users is becoming very popular~\cite{Bassily_SPM_2013}. Also due to its relative difficulty as compared to source based broadcast, because of the possibility of information interception in both the hops, significant research attention is being paid in this regard recently~\cite{amitav_TCST_2014}. 

The authors in \cite{LLai_TIT_2008} proposed several cooperation strategies for secrecy enhancement in single carrier communication systems. While considering four single antenna half duplex nodes, \cite{Deng_TIFS_2015} investigated the role to be played by the relay to maximize  ergodic secrecy rate. For a similar setting, \cite{Hmwang_TC_2015} considered the outage constrained secrecy throughput maximization problem. 

{\color{black}
In an amplify and forward (AF) relay assisted system without availability of direct source-destination link, a time division based protocol was proposed in~\cite{HXu_TIFS_2017} using one of the users as the helper node for secure communication to untrusted users.  
In another related work~\cite{Amabrouk_PIMRC_2016}, time division based relay and user selection scheme was studied to improve secrecy of a cooperative AF relay network, assuming availability of direct link.}  
With multi-antenna nodes, \cite{Derrick_TWC_2011} investigated multiuser resource allocation for decode and forward (DF) relay assisted system without direct link, in the presence of single eavesdropper. 
The authors in \cite{RSaini_CL_2016} considered resource allocation problem for a DF relay assisted orthogonal frequency division multiple access (OFDMA) system with multiple untrusted users. 

Assuming the availability of direct link, the optimal power allocation and transmission mode selection for DF relay-assisted secure communication was considered in \cite{Jeong_TSP_2011}. Observing that \textit{strategies for a single source-destination pair with joint transmit power budget for source and relay cannot be extended for an  untrusted users' model with individual power budgets}, we intend to investigate whether utilizing a relay is always useful in multiuser secure OFDMA system. 
 


The key contributions of this letter are four fold. {\color{black}\emph{Firstly}, we present a generalized secure rate definition for DF relay assisted secure OFDMA system with the availability of direct link, while considering the possibility of tapping in both the hops.}  \emph{Secondly}, observing that each subcarrier can be utilized in direct communication (DC) mode, we identify the conditions for using a subcarrier in relayed communication (RC) mode, and obtain optimal subcarrier allocation policies for both modes.  
{\color{black}\emph{Thirdly}, noting that a set of subcarriers can be used in both the  modes, we find optimal mode selection strategy resulting in higher secure rate over such subcarriers. \emph{Finally}, asymptotically optimal and suboptimal mode selection schemes, that are independent of power allocation, are derived.} 
\emph{To the best of our knowledge, it is the first work studying utility of a DF relay in secure OFDMA system with untrusted users.}

\section{System model}
{\color{black}Downlink of a trusted DF relay $\mathcal{R}$ assisted secure OFDMA system, with  source $\mathcal{S}$, and $M$ untrusted users is considered. Untrusted users is a hostile scenario, where each user behaves as a potential eavesdropper for others. For each $\mathcal{U}_m$ there are effectively $M-1$ eavesdroppers, and the one having maximum signal-to-noise ratio (SNR) is called \textit{equivalent eavesdropper}. Apart from the direct $(\mathcal{S}-\mathcal{U}_m)$ link, there exists a two hop  $(\mathcal{S}-\mathcal{R})$ and  $(\mathcal{R}-\mathcal{U}_m)$ link for information transfer to $\mathcal{U}_m$.  
} 

\emph{Assumptions}: All nodes are equipped with single antenna, and $\mathcal{R}$ operates in two hop half duplex DF mode \cite{Derrick_TWC_2011, Jeong_TSP_2011}. All  subcarriers on $\mathcal{S}-\mathcal{R}$, $\mathcal{S}-\mathcal{U}_m$,  $\mathcal{R}-\mathcal{U}_m$ links are assumed to follow quasi-static Rayleigh fading.  Perfect channel state information over all links is available at $\mathcal{S}$ \cite{Xiaowei_TIFS_2011, Jeong_TSP_2011, Derrick_TWC_2011}. 
Users are capable of utilizing maximum ratio combining (MRC)\cite{Jeong_TSP_2011}.

\section{Proposed Secure Rate Definition}\label{sec:secure_rate_definition}
Before introducing secure rate definition in an untrusted user scenario with two tapping, we first discuss rate definitions in classical co-operative communication. 
Let us denote the rate achieved by user  $\mathcal{U}_m$ over subcarrier $n$ in DC and RC mode as $R_{n}^m|_{DC}$ and $R_{n}^m|_{RC}$, respectively. 
{\color{black}With $\mathcal{S}$ utilizing optimum transmission mode for achieving maximum secure rate,} the effective rate $R_{n}^m$ is given by  
$R_{n}^m = \max \left \{ R_{n}^m|_{DC}, R_{n}^m|_{RC} \right \}$.

\subsection{Rate Definitions in Classical Co-operative Communication}
Let $R_{n}^{sm}$, $R_{n}^{sr}$, and $R_{n}^{srm}$, respectively, denote the rates of $\mathcal{U}_m$ for $\mathcal{S}-\mathcal{U}_m$, $\mathcal{S}-\mathcal{R}$, and   $\mathcal{S}-\mathcal{R}-\mathcal{U}_m$ links over subcarrier $n$. Here $R_{n}^{srm}$ denotes the rate of $\mathcal{U}_m$ due to MRC of signals from $\mathcal{S}$ and $\mathcal{R}$. The rates of $\mathcal{U}_m$ in DC and RC modes are:
\begin{equation}\label{rate_definition_1}
{\color{black}R_{n}^m|_{DC} = R_{n}^{sm}};\quad R_{n}^m|_{RC} = \left(1/2\right) \min \left \{ R_{n}^{sr}, R_{n}^{srm} \right \}.
\end{equation}
The factor $\frac{1}{2}$ in {\color{black}$R_{n}^m|_{RC}$} arises due to the half duplex protocol. 
Thus, 
$R_{n}^m = \frac{1}{2} \max \left \{ {\color{black}2R_{n}^{sm}}, \min \left \{ R_{n}^{sr}, R_{n}^{srm} \right \} \right\}$.

Let $P_{n}^s$ and $P_{n}^r$, respectively, denote source and relay power over subcarrier $n$. The channel gain of $i$--$j$ link over subcarrier $n$ is denoted by $\gamma_{n}^{ij}$ where $i \in \{ s, r \}$ and $j \in \{ r, 1, 2, \cdots M \}$. The rates of $\mathcal{S}-\mathcal{U}_m$,    $\mathcal{S}-\mathcal{R}$, and $\mathcal{S}-\mathcal{R}-\mathcal{U}_m$ links are respectively given by $R_{n}^{sm} = \log_2 \left( 1+ {P_{n}^s\gamma_{n}^{sm}}/{\sigma^2}\right)$, $R_{n}^{sr} = \log_2 \Big( 1+$ $\frac{P_{n}^s\gamma_{n}^{sr}}{\sigma^2}\Big)$, and $R_{n}^{srm} = \log_2 \left( 1+ \frac{P_{n}^s\gamma_{n}^{sm} + P_{n}^r\gamma_{n}^{rm}}{\sigma^2}\right)$. After some simplifications the rate $R_{n}^m$ 
can be restated as
\begin{align}\label{simplified_rate_def}
\textstyle{\color{black}
R_{n}^m = \frac{1}{2}
\begin{cases}\textstyle
R_{n}^{sr} & \text{if $2R_{n}^{sm} \leq R_{n}^{sr} < R_{n}^{srm}$ }\\
R_{n}^{srm} & \text{if $2R_{n}^{sm} \leq R_{n}^{srm} \leq R_{n}^{sr}$ }\\
2R_{n}^{sm} & \text{otherwise}.
\end{cases}
} 
\end{align}

{\color{black}
$2R_{n}^{sm} \leq R_{n}^{sr}$ can be simplified as $\gamma_{n}^{sr} \geq \gamma_{n}^{sm} a_{n}^m$ where $a_n^m = \left(2+\frac{P_{n}^s\gamma_{n}^{sm}}{\sigma^2}\right)$, which upper bounds $P_{n}^s$ as $P_{n}^s \leq P_{n_u}^{sm} \triangleq \frac{(\gamma_{n}^{sr}-2\gamma_{n}^{sm})\sigma^2}{(\gamma_{n}^{sm})^2}$. $2R_{n}^{sm} \leq R_{n}^{srm}$ leads to $P_{n}^r  \geq P_{n_l}^{rm} \triangleq P_{n}^s \frac{\gamma_{n}^{sm}}{\gamma_{n}^{rm}} \left( 1 + \frac{P_{n}^s\gamma_{n}^{sm}}{\sigma^2} \right)$. Thus, if $P_{n}^s$ is below a certain threshold, and $P_{n}^r$ is above a certain threshold, RC mode can be used, otherwise DC mode is a better option.} 
$R_{n}^{sr}<R_{n}^{srm}$ leads to $\frac{P_{n}^r}{P_{n}^s}> \frac{\gamma_{n}^{sr}-\gamma_{n}^{sm}}{\gamma_{n}^{rm}} \triangleq \Delta_n^m$, where $\Delta_n^m$ is referred as relay versus source power (RSP) ratio. Thus,
  $R_{n}^m$ \eqref{simplified_rate_def} can be simplified as
\begin{align}\label{simplified_rate_def_1}
{\color{black}
\hspace{-2 mm} R_{n}^m \hspace{-1 mm} = \hspace{-1 mm}
\begin{cases}
\hspace{-1 mm}\frac{1}{2} R_{n}^{sr} & \hspace{-3 mm} \text{if \hspace{-1 mm}$\gamma_{n}^{sr}\geq\gamma_{n}^{sm}a_n^m, P_{n}^r \geq \max \{ P_{n_l}^{rm},  P_{n}^s\Delta_n^m$\}} \\ 
\hspace{-1 mm}\frac{1}{2} R_{n}^{srm} & \hspace{-3 mm} \text{if \hspace{-1 mm} $\gamma_{n}^{sr}\geq\gamma_{n}^{sm}a_n^m, P_{n}^s\Delta_n^m \geq P_{n}^r \geq P_{n_l}^{rm}$ } \\  
\hspace{-1 mm}R_{n}^{sm} & \hspace{-3 mm} \text{otherwise}. \hspace{-3 mm}
\end{cases} 
}
\end{align}


\begin{remark}
\textit{From \eqref{simplified_rate_def_1}, we note that, if $P_{n}^r \leq P_{n}^s\Delta_n^m$, MRC link $\mathcal{S}-\mathcal{R}-\mathcal{U}_m$ is the bottleneck compared to  $\mathcal{S}-\mathcal{R}$ link, and the rate is $R_{n}^{srm}$. As $\gamma_{n}^{sr} \geq \gamma_{n}^{sm}$, MRC link remains as the bottleneck even for increased $P_{n}^s$. This rate in RC mode can be improved by increasing $P_{n}^r$ till $R_{n}^{sr}=R_{n}^{srm}$, after which $\mathcal{S}-\mathcal{R}$ link becomes the bottleneck. Thus, maximum rate in RC mode  is achieved when $R_{n}^{sr}=R_{n}^{srm}$, i.e., $P_{n}^r = P_{n}^s\Delta_n^m$.}
\end{remark}

\subsection{Incompleteness of Classical Rate Definition}\label{subsec_completeness}
The rate definition of $R_{n}^m|_{RC}$ in RC mode is based on an implicit assumption that $P_n^r>0$. When $P_n^r=0$, $R_{n}^{srm} = R_{n}^{sm}$, and $R_{n}^m|_{RC}=\frac{1}{2} \min \left \{ R_{n}^{sr}, R_{n}^{sm} \right \}$ which is positive for $P_n^s>0$. But this has no physical significance as the decoded information at $\mathcal{R}$ is not forwarded to $\mathcal{U}_m$. Ideally, $P_n^r=0$ should indicate that $R_{n}^m|_{RC} = 0$, such that $R_{n}^m = R_{n}^m|_{DC}$. 

The proposed rate definition is complete as it takes care of this gap. With $P_n^r=0$ and $R_{n}^{srm} = R_{n}^{sm}$, the definition  gets simplified to {\color{black}$R_{n}^m = \frac{1}{2} \max \left \{ 2R_{n}^{sm}, \min \left \{ R_{n}^{sr}, R_{n}^{sm} \right \} \right \}$. Thus, with $P_n^r=0$, when either $R_{n}^{sr} < R_{n}^{sm}$ or $R_{n}^{sr} \geq R_{n}^{sm}$, rate $R_{n}^m = 2R_{n}^{sm} = R_{n}^m|_{DC}$, i.e., subcarrier is used in DC mode.} 


\subsection{Secure Rate Definition}\label{sec:model}
The secure rate $R_{s_{n}}^m$ of $\mathcal{U}_m$ over a subcarrier $n$ is the difference of  rate $R_{n}^m$ of $\mathcal{U}_m$ and rate $R_{n}^e$ of the equivalent eavesdropper $\mathcal{U}_e$ \cite{Xiaowei_TIFS_2011}. Mathematically, $R_{s_{n}}^m$ is given by
\begin{align}\label{rate_definition_3}
\textstyle R_{s_{n}}^m = \left[ R_{n}^m - R_{n}^e \right]^+ = \Big[ R_{n}^m - \max\limits_{o \in \{1,2,\cdots M\} \setminus m} R_{n}^o\Big]^+
\end{align}
where $x^+ = \max\{0,x\}$. The definition in \eqref{rate_definition_3} considers tapping in both slots. 
{\em Further, in contrast to the secure rate definition used in \cite{Hmwang_TC_2015} and \cite{Derrick_TWC_2011}, which did not consider direct link availability, the proposed definition is a generalized one.}

\section{Subcarrier Allocation Policy}
Now, we discuss the conditions for achieving positive secure rate by $\mathcal{U}_m$ over a subcarrier $n$. From \eqref{simplified_rate_def_1}, a subcarrier can be utilized in either DC or RC mode. In DC mode, the required condition is  $R_{n}^{sm}>R_{n}^{se}$ which can be simplified as $\gamma_{n}^{sm}>\gamma_{n}^{se}$. 
With $\pi_{n_{DC}}^m$ denoting the subcarrier allocation variable in DC mode, the subcarrier allocation policy can be stated as
\begin{align}\label{subcarrier_alloc_relay_dc}
\pi_{n_{DC}}^m=
\begin{cases}
1 & \text{if $m = \arg \max \limits_{o \in \{1,2,\cdots M\}} \gamma_{n}^{so}$}\\
0 & \text{otherwise.}
\end{cases} 
\end{align}
Positive secure rate conditions for RC mode are given below. 

\begin{proposition}\label{proposition_rate_positivity}
\textit{{\color{black}$\mathcal{U}_m$ can use a subcarrier $n$ in RC mode if: (i) $\gamma_{n}^{sr} > \max \{ \gamma_{n}^{so} a_n^o$ \}, (ii) $P_{n}^r > \max \{ P_{n_l}^{ro} \} $}  (iii) $P_{n}^r \leq P_{n}^s \Delta_n^m$, and (iv) $\Delta_n^m = \min \{ \Delta_n^o \} \text{ } \forall o \in \{1,2,\cdots M\}$.
}
\end{proposition}

\begin{IEEEproof}
The conditions for activating RC mode over subcarrier $n$ are {\color{black}$\gamma_{n}^{sr} \geq \gamma_{n}^{sm} a_n^m$ and $P_{n}^r \geq P_{n_l}^{rm}$ (cf. \eqref{simplified_rate_def_1}). 
Its generalization for $M$ users leads to the first and the second  conditions: $\gamma_{n}^{sr} \geq \max \limits_{ o \in \{1,2,\cdots M\}} \gamma_{n}^{so} a_n^o$ and $P_{n}^r \geq \max \limits_{ o \in \{1,2,\cdots M\}} P_{n_l}^{ro}$.}

Let $\Delta_n^e$ denote RSP ratio (cf. \eqref{simplified_rate_def_1}) for  $\mathcal{U}_e$ over subcarrier $n$. If $P_{n}^r >  P_{n}^s \Delta_n^m$, rate of $\mathcal{U}_m$ is $R_{n}^{sr}$. The rate of $\mathcal{U}_e$ is either $R_{n}^{sr}$ when $P_{n}^r >  P_{n}^s \Delta_n^e$, or $R_{n}^{sre}$ otherwise. In the first case the secure rate is zero, while in the second case $R_{s_{n}}^m = R_{n}^{sr}-R_{n}^{sre}$ = $\frac{1}{2}\left\{ \log_2 \left( \frac{\sigma^2+P_{n}^s\gamma_{n}^{sr}}{\sigma^2+P_{n}^s\gamma_{n}^{se} + P_{n}^r\gamma_{n}^{re}}\right) \right\}$, which is a decreasing function of $P_{n}^r$,  enforcing $P_{n}^r=0$, i.e., DC mode (cf. Section \ref{subsec_completeness}). Thus, for  RC mode $P_{n}^r \leq P_{n}^s \Delta_n^m$ which is second condition. Lastly, we prove the third condition $\Delta_n^e > \Delta_n^m$ by contradiction that if $\Delta_n^e \leq \Delta_n^m$ then positive secure rate cannot be achieved. The condition $\Delta_n^e \leq \Delta_n^m$ can be restated as
\begin{align}\label{prop_proof_1}
\textstyle\frac{\gamma_{n}^{sr}-\gamma_{n}^{se}}{\gamma_{n}^{re}} \leq
\frac{\gamma_{n}^{sr}-\gamma_{n}^{sm}}{\gamma_{n}^{rm}}.
\end{align}

Simplifying $\gamma_{n}^{sr}$ from the definition of $\Delta_n^e$, we get $\gamma_{n}^{sr} = \gamma_{n}^{se}+\Delta_n^e \gamma_{n}^{re}$. Substituting $\Delta_n^e$ in \eqref{prop_proof_1}, we obtain $\gamma_{n}^{sr} 
\geq  \gamma_{n}^{sm}+\Delta_n^e \gamma_{n}^{rm}$. Substituting $\gamma_{n}^{sr}$ results in  $\gamma_{n}^{se}+\Delta_n^e \gamma_{n}^{re} \geq \gamma_{n}^{sm}+\Delta_n^e \gamma_{n}^{rm}$. Multiplying both the sides with $P_{n}^s$, and substituting $P_{n}^s \Delta_n^e$ as $P_{n}^r$, we get  $P_{n}^s\gamma_{n}^{se}+P_{n}^r\gamma_{n}^{re}  \geq P_{n}^s\gamma_{n}^{sm}+P_{n}^r\gamma_{n}^{rm}$, which will lead to zero secure rate as $R_{n}^e \geq R_{n}^m$. Thus, to achieve positive secure rate $\Delta_n^e > \Delta_n^m$. Under this condition, the rates of $\mathcal{U}_m$ and $\mathcal{U}_e$ are given as $R_{n}^{srm}$ and $R_{n}^{sre}$, respectively, and the secure rate definition in \eqref{rate_definition_3} gets simplified to
\begin{align}\label{rate_defintion_MRC}
\textstyle R_{s_{n}}^m = \frac{1}{2} \log_2 \left( \frac{\sigma^2+P_{n}^s\gamma_{n}^{sm} + P_{n}^r\gamma_{n}^{rm}}{\sigma^2+P_{n}^s\gamma_{n}^{se} + P_{n}^r\gamma_{n}^{re}}\right).
\end{align}
 
The condition $\Delta_n^e>\Delta_n^m$ must be satisfied for all possible $\mathcal{U}_e$. Thus $\mathcal{U}_m$ has to be chosen for having minimum ratio  $ \Delta_n^o  \forall o \in \{1,2,\cdots M\}$. With $\pi_{n_{RC}}^m$ as subcarrier allocation variable in RC mode, optimal subcarrier allocation policy is
\begin{align}\label{subcarrier_alloc_relay_rc}
\textstyle\pi_{n_{RC}}^m=
\begin{cases}\textstyle
1 & \text{if $m = \arg \min \limits_{o \in \{1,2,. M\}} \left( \Delta_n^o \triangleq \frac{\gamma_{n}^{sr} - \gamma_{n}^{so}}{\gamma_{n}^{ro}}  \right) $}\\
0 & \text{otherwise.}
\end{cases} 
\end{align}

After sorting RSP ratios ($\Delta_n^o$) over a subcarrier in ascending order, the user having the minimum value is $\mathcal{U}_m$, and the one having just next better value is the corresponding $\mathcal{U}_e$. 
\end{IEEEproof}

{\em
Physical Interpretation of \eqref{subcarrier_alloc_relay_rc}: From \eqref{simplified_rate_def_1}, RSP ratio $\Delta_n^o = \frac{\gamma_{n}^{sr} - \gamma_{n}^{so}}{\gamma_{n}^{ro}}$ is the factor by which $P_{n}^r$ should be provided for a fixed $P_{n}^s$ to achieve the same SNR over $\mathcal{S}-\mathcal{R}$ and $\mathcal{S}-\mathcal{R}-\mathcal{U}_m$ links. So a user having a lower ratio will require lower $P_{n}^r$ to achieve maximum secure rate. Thus, once a  user is chosen with minimum value of the ratio as the main user $\mathcal{U}_m$, then for any other user $\mathcal{U}_e$ having higher value of RSP ratio, its $\mathcal{R}-\mathcal{U}_e$ link becomes the bottleneck link (as it requires higher $P_{n}^r$ to become equal to the $\mathcal{S}-\mathcal{R}$ link) and its rate will be lower than that of the main user. Thus, allocation in \eqref{subcarrier_alloc_relay_rc} always leads to positive secure rate over a subcarrier in RC mode. 
}

\begin{remark}
\textit{Due to the possibility of tapping in the first slot, the condition $\gamma_{n}^{sm}>\gamma_{n}^{se}$ (cf. \eqref{subcarrier_alloc_relay_dc}) must be satisfied in RC mode as well. So, the main user in RC mode (cf. \eqref{subcarrier_alloc_relay_rc}) also satisfies positive secure rate requirement for DC mode (cf. \eqref{subcarrier_alloc_relay_dc}) .}
\end{remark}

\begin{remark}
\textit{In case the same user is selected as  main user through the policies \eqref{subcarrier_alloc_relay_dc} and \eqref{subcarrier_alloc_relay_rc}, that subcarrier satisfies positive secure rate requirement for both DC and RC modes. However, corresponding eavesdroppers in the two modes can be different. With $\mathcal{U}_e$ and $\mathcal{U}_{e'}$  respectively denoting eavesdroppers in RC and DC modes over $n$, from  \eqref{subcarrier_alloc_relay_dc}: $\gamma_n^{se'} \geq \gamma_n^{se}.$} \end{remark}

\section{Utility of Relay: RC versus DC Mode Selection }
To highlight the utility of relay, here we present the conditions for enhanced performance of RC mode over DC mode. 
Thus, we intend to derive conditions for $R_{s_{n}}^m|_{RC}>R_{s_{n}}^m|_{DC}$. Consider the general case where the eavesdroppers of user $\mathcal{U}_m$ are different in RC mode ($\mathcal{U}_e$) and DC mode ($\mathcal{U}_{e'}$). The condition can be simply stated as $(R_{n}^m-R_{n}^e)|_{RC}  > (R_{n}^m-R_{n}^{e'})|_{DC}$. 
\begin{eqnarray}\label{rate_comparison}
\textstyle{\color{black}\frac{1}{2}\log_2 \left( \frac{\sigma^2+P_{n}^s\gamma_{n}^{sm} + P_{n}^r\gamma_{n}^{rm}}{\sigma^2+P_{n}^s\gamma_{n}^{se} + P_{n}^r\gamma_{n}^{re}}\right) >\log_2 \left( \frac{\sigma^2+P_{n}^s\gamma_{n}^{sm}}{\sigma^2+P_{n}^s\gamma_{n}^{se'}}\right).}
\end{eqnarray}	
{\color{black}Using energy efficient solution $P_{n}^r = P_{n}^s \Delta_n^m$ \cite{RSaini_CL_2016}, the resulting condition gets simplified as}:
\begin{eqnarray}\label{rate_comparison_1}
\textstyle{\color{black}\frac{\gamma_{n}^{rm}}{\gamma_{n}^{re}} > \rho \triangleq \frac{(\gamma_{n}^{sr}-\gamma_{n}^{sm})(\sigma^2+P_{n}^s\gamma_{n}^{sm})^2}{\rho_{den}}.}
\end{eqnarray}
{\color{black}
where $\rho_{den} = \gamma_{n}^{sr}(\sigma^2+P_{n}^s\gamma_{n}^{se'})^2-\gamma_{n}^{se}(\sigma^2+P_{n}^s\gamma_{n}^{sm})^2-\sigma^2\{(\sigma^2+P_{n}^s\gamma_{n}^{se'})+(\sigma^2+P_{n}^s\gamma_{n}^{sm})\}(\gamma_{n}^{sm}-\gamma_{n}^{se'})$.
$\rho<0$ indicates exclusive DC mode.
Next, we discuss mode selection under asymptotic conditions, and with and without known $P_n^s$.}

\subsection{Asymptotically Optimal Mode Selection Policy}\label{low_high_snr_bounds}
At low SNR regime, \eqref{rate_comparison} can be simplified using approximation $\log(1+x) \approx x, \forall x \ll 1$, and $P_{n}^r = P_{n}^s \Delta_n^m$, as:
\begin{align}\label{low_snr_policy}
\textstyle{\color{black}\frac{\gamma_{n}^{rm}}{\gamma_{n}^{re}} > \rho_{l} \triangleq \frac{\gamma_{n}^{sr}-\gamma_{n}^{sm}}{\gamma_{n}^{sr}-2(\gamma_{n}^{sm}-\gamma_{n}^{se'})-\gamma_{n}^{se}}}. 
\end{align}

Under high SNR scenario, using the approximation $\log(1+x) \approx \log(x), \forall x \gg 1$, the condition in \eqref{rate_comparison} gets simplified to:  
\begin{align}\label{high_snr_policy}
\textstyle{\color{black}
\frac{\gamma_{n}^{rm}}{\gamma_{n}^{re}}>\rho_{h} \triangleq \frac{(\gamma_{n}^{sr}-\gamma_{n}^{sm})(\gamma_{n}^{sm})^2}{\gamma_{n}^{sr}(\gamma_{n}^{se'})^2-\gamma_{n}^{se}(\gamma_{n}^{sm})^2}}
\end{align}



\subsection{Optimal Mode Selection for given Power Allocation}
{\color{black}
First, we show that $\rho_l<\rho$. Thus, if $\frac{\gamma_{n}^{rm}}{\gamma_{n}^{re}}<\rho_l$, the subcarrier has to be used exclusively in DC mode. Referring \eqref{rate_comparison_1} and \eqref{low_snr_policy}, condition $\rho_l<\rho$ can be stated as:
$\gamma_{n}^{sr}-2(\gamma_{n}^{sm}-\gamma_{n}^{se'})-\gamma_{n}^{se}>\frac{\rho_{den}}{(\sigma^2+P_n^s\gamma_{n}^{sm})^2}$.
Substituting $\rho_{den}$, this gets simplified as:
\begin{align}
\textstyle&\gamma_{n}^{sr}-2(\gamma_{n}^{sm}-\gamma_{n}^{se'})-\gamma_{n}^{se} > \textstyle\gamma_{n}^{sr} \Big( \frac{\sigma^2+P_n^s\gamma_{n}^{se'}}{\sigma^2+P_n^s\gamma_{n}^{sm}} \Big)^2 \nonumber \\
&\textstyle \quad - \sigma^2 \Big( \frac{(\sigma^2+P_n^s\gamma_{n}^{sm})+(\sigma^2+P_n^s\gamma_{n}^{se'})}{(\sigma^2+P_n^s\gamma_{n}^{sm})^2}\Big) - \gamma_{n}^{se}.
\end{align}
After arranging terms and some simplification steps, we obtain  
$(\gamma_{n}^{sm}-\gamma_{n}^{se'})\left[\frac{(2\sigma^2+P_n^s(\gamma_{n}^{sm}+\gamma_{n}^{se'}))(\sigma^2+P_n^s\gamma_{n}^{sr})}{(\sigma^2+P_n^s\gamma_{n}^{sm})^2}-2\right]>0.$
With $P_n^s>0$ and $(\gamma_{n}^{sm}-\gamma_{n}^{se'})>0$, it gets reduced to
$(\gamma_{n}^{sr}-2\gamma_{n}^{sm})(2\sigma^2+P_n^s\gamma_{n}^{sm}) + \sigma^2 (\gamma_{n}^{sm}+\gamma_{n}^{se'})+P_n^s \gamma_{n}^{sr} \gamma_{n}^{se'}>0.$
Observing that $\gamma_{n}^{sr}>2\gamma_{n}^{sm}$, the above condition always holds.

Similarly, we prove that $\rho_h>\rho$, such that if $\frac{\gamma_{n}^{rm}}{\gamma_{n}^{re}}>\rho_h$, the subcarrier should be in RC mode exclusively. The equivalent condition for $\rho_h>\rho$ can be stated as (cf. \eqref{rate_comparison_1} and \eqref{high_snr_policy}):
\begin{eqnarray}\textstyle
\textstyle\Big( \frac{\gamma_{n}^{se'}}{\gamma_{n}^{sm}} \Big) ^2 < \Big( \frac{\sigma^2+P_n^s\gamma_{n}^{se'}}{\sigma^2+P_n^s\gamma_{n}^{sm}} \Big)^2 - \frac{\sigma^2}{\gamma_{n}^{sr}} \Big( \frac{(\sigma^2+P_n^s\gamma_{n}^{sm})+(\sigma^2+P_n^s\gamma_{n}^{se'})}{(\sigma^2+P_n^s\gamma_{n}^{sm})^2}\Big)
\end{eqnarray}
After arranging the terms, this condition gets simplified as $\sigma^2(\gamma_{n}^{sm}-\gamma_{n}^{se'})\gamma_{n}^{sr}\left[\sigma^2(\gamma_{n}^{sm}+\gamma_{n}^{se'})+2P_n^s\gamma_{n}^{sm}\gamma_{n}^{se'}\right] > \sigma^2(\gamma_{n}^{sm}-\gamma_{n}^{se'})(\gamma_{n}^{sm})^2\left[2\sigma^2 +P_n^s(\gamma_{n}^{sm}+\gamma_{n}^{se'})\right]$. 
With $\gamma_{n}^{sm}>\gamma_{n}^{se'}$, and rearranging the terms, the condition gets reduced to $\sigma^2\gamma_{n}^{sm}\left( \gamma_{n}^{sr}-2\gamma_{n}^{sm}-\frac{P_n^s(\gamma_{n}^{sm})^2}{\sigma^2} \right) + \sigma^2 \gamma_{n}^{sr}\gamma_{n}^{se'} + P_n^s\gamma_{n}^{sm}\gamma_{n}^{se'}(2\gamma_{n}^{sr}-\gamma_{n}^{sm})>0$, which is always true as $P_n^s<P_{n_u}^{sm}$ and $\gamma_{n}^{sr}>\gamma_{n}^{sm}$. The complete mode selection policy with known power allocation can be summarized as: 
\begin{eqnarray}\label{optimal_mode_selection}
\textstyle  \frac{\gamma_{n}^{rm}}{\gamma_{n}^{re}}\textstyle  \begin{cases}\textstyle
  >\rho_h & \text{$R_{s_{n}}^m|_{RC} > R_{s_{n}}^m|_{DC}$ \textbf{Exclusive RC}}\\
  \in\left[\rho_l,\rho_h\right] & 
      \textbf{RDC}      
      \begin{cases}\textstyle
      P_{n}^s<P_{n_{th}}^s & \hspace{-1mm}\textbf{RC} \\
      P_{n}^s \geq P_{n_{th}}^s & \hspace{-1mm}\textbf{DC} \\
	  \end{cases}	      
      \\
  <\rho_l & \text{$R_{s_{n}}^m|_{RC} < R_{s_{n}}^m|_{DC}$ \textbf{Exclusive DC}}
  \end{cases}\hspace{-2.5mm}
\end{eqnarray}
where $P_{n_{th}}^s$ is positive root of quadratic obtained from \eqref{rate_comparison_1}.

{\em Physical Interpretation of \eqref{optimal_mode_selection}: Secure rate improvement with $P_n^r$ depends on relative gain $\frac{\gamma_{n}^{rm}}{\gamma_{n}^{re}}$. 
 In low SNR case, all RDC mode subcarriers are in RC mode as $P_n^s<P_{n_{th}}^s$. Thus, if $\frac{\gamma_{n}^{rm}}{\gamma_{n}^{re}}<\rho_l$, the subcarrier is in DC mode,  otherwise it can be in RC mode. In high SNR case, with $P_n^s>P_{n_{th}}^s$, all RDC mode subcarriers switch to DC mode. Only those subcarriers which have $\frac{\gamma_{n}^{rm}}{\gamma_{n}^{re}}>\rho_h$ are in RC mode, rest are in DC mode. 
}
}

\subsection{Sub-optimal Mode Selection Policy}
We now propose a suboptimal mode selection strategy that does not require explicit knowledge of $P_n^s$. 
Let us introduce a term `satisfaction level' $\alpha$ which is considered as the minimum acceptable SNR level over a subcarrier, i.e., $\frac{P_n^s \gamma_{n}^{sm}}{\sigma^2}>\alpha \,\forall n$. As higher value of $\alpha$ requires higher source power on each subcarrier, it can be considered as an abstraction parameter mapping minimum supported SNR to source power budget. 

\begin{remark} 
\textit{As $\gamma_{n}^{sr}>\gamma_{n}^{sm}$ and $P_n^s \gamma_{n}^{sm}+P_n^r \gamma_{n}^{rm}>P_n^s \gamma_{n}^{sm}$, $\frac{P_n^s \gamma_{n}^{sm}}{\sigma^2}>\alpha$ is enough to  ensure successful communication.}
\end{remark}

\textcolor{black}
{To have a higher secure rate in RC mode than in DC mode $P_{n_{th}}^s>P_{n}^s>\frac{\sigma^2\alpha}{\gamma_{n}^{sm}}$. Substituting $P_{n}^s$, we have: ${\color{black}
\frac{\gamma_{n}^{rm}}{\gamma_{n}^{re}} > \rho_{\alpha} \triangleq \frac{\alpha_{num}}{\alpha_{den}}},$ where $\alpha_{num} = (\gamma_{n}^{sr}-\gamma_{n}^{sm})(1+\alpha)^2$, and $\alpha_{den} = \gamma_n^{sr}(1 + \alpha \frac{\gamma_{n}^{se'}}{\gamma_{n}^{sm}})^2-\gamma_n^{se}(1 + \alpha)^2 - (\gamma_n^{sm}-\gamma_n^{se'})\{(1+\alpha)+ (1+\alpha \frac{\gamma_{n}^{se'}}{\gamma_{n}^{sm}}) \}$. 
For the limiting cases $\alpha \to 0$ and $\alpha \to \infty$, $\rho_{\alpha}$ respectively tends to the low and high SNR bounds $\rho_l$ and $\rho_u$ on $\frac{\gamma_{n}^{rm}}{\gamma_{n}^{re}}$ discussed in Section \ref{low_high_snr_bounds}}. This corroborates our reasoning behind $\alpha$ being a measure of source power budget. 

\section{Numerical Results}\color{black}
The downlink of an OFDMA system is considered with $N=64$ subcarriers which are assumed to experience quasi-static Rayleigh fading with path loss exponent $=3$. We study performance variation with relay position, secure rate improvement due to optimal mode selection and utility regions.

\begin{figure}[!htb]
\centering
\includegraphics[scale=.5]{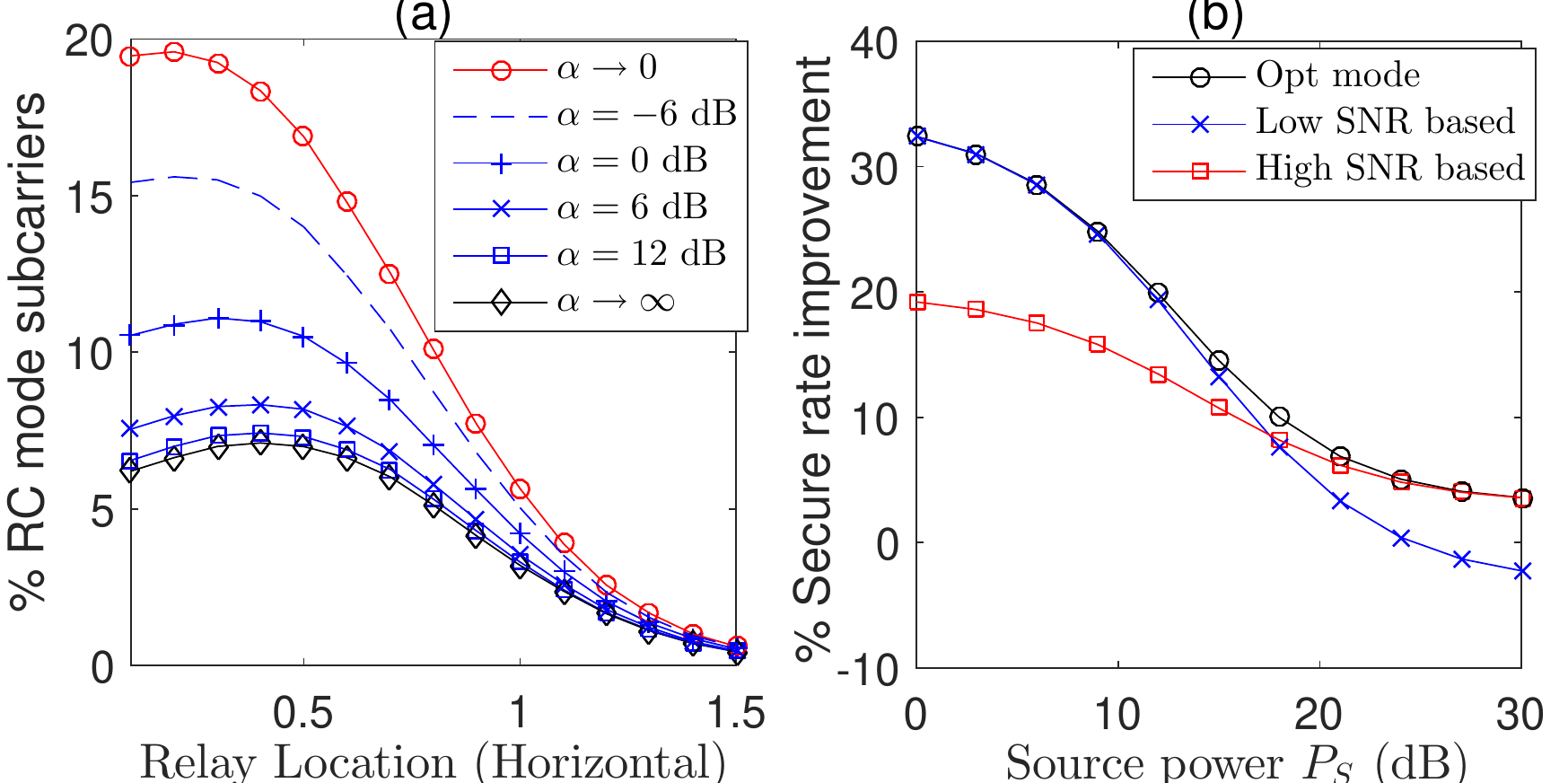} 
\caption{(a) Performance with horizontal variation in relay  position, (b) Secure rate improvement through mode selection.}
\label{fig:relay_position}
\end{figure}

Fig. \ref{fig:relay_position}(a) presents the effect of relay placement on its utility in improving the secure rate. Considering DC mode as a benchmark, improvement in system performance is presented by plotting the percentage of subcarriers that have higher rate in RC mode. Assuming $\mathcal{S}$ to be located at $(0,0)$ and $M=8$ users randomly distributed inside a unit square centered at $(2,0)$, position of $\mathcal{R}$ is varied along a horizontal line $(x_r, 0)$ with $0.1 \leq x_r \leq 1.5$.
Note that, $\mathcal{R}$ should placed closer to $\mathcal{S}$, to have $\gamma_n^{sr} \geq \gamma_n^{sm}a_n^m$ and stand against DC mode. Source power budget variation is captured by varying $\alpha$. Even though optimal relay location $x_r^*$ increases with $\alpha$, it is still in the left half, i.e., $x_r^*<0.5$, for the considered system. Note that with increased $\alpha$ percentage of RC mode subcarriers reduces as more and more RDC mode subcarrier switches to DC mode.

Considering equal power allocation, rate improvement achieved by optimal mode selection compared to static DC mode is plotted in Fig \ref{fig:relay_position}(b). Following the observation from Fig \ref{fig:relay_position}(a), relay is placed at $(0.5,0)$. Performance of low and high SNR based policies have been plotted to highlight efficacy of optimal policy. Rate improvement reduces with increasing $P_S$ as all RDC subcarriers move to DC mode. At higher $P_S$, negative improvement is observed in low SNR based policy because RDC subcarriers which could have achieved higher rate in DC mode are pushed to RC mode. 

\begin{figure}[!htb]
\centering
\includegraphics[scale=.5]{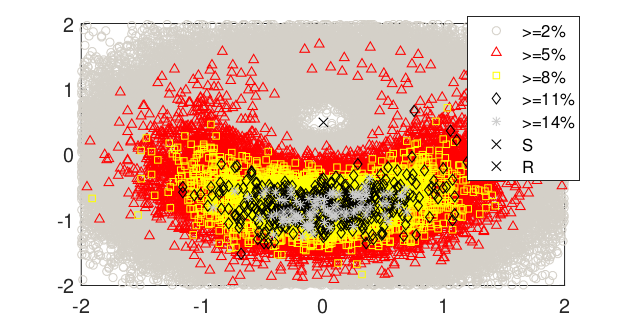} 
\caption{Relay utility regions.}
\label{fig:relay_utility_region}
\end{figure} 

Fig \ref{fig:relay_utility_region} presents spatial utility of relay where users' locations on a 2-D Euclidean plane are plotted after categorizing them according to the percentage of RC mode subcarriers. Assuming users to be located randomly in a $4 \times 4$ square centered around $(0,0)$, $\mathcal{S}$ and $\mathcal{R}$ are considered to be located at $(0,0.5)$ and $(0, -0.5)$, respectively. \textit{Note that the best utility is around relay where more than $14\%$ subcarriers are benefited by RC mode.} Due to direct link availability, decreasing trend of percentage RC mode subcarriers with distance is not symmetric.
 
\color{black}

\section{Conclusion} 
Considering two slot tapping, this paper presents a generalized secure rate definition.  After identifying conditions for RC mode, optimal subcarrier allocation policies for both RC and DC modes are obtained. A subcarrier can be used either in exclusive DC mode, in exclusive RC mode, or in RDC mode. \textcolor{black}{Identifying that optimal mode selection policy for RDC mode subcarriers is integrated with power allocation, an $\alpha$ based suboptimal policy is discussed, which asymptotically matches with the optimal policy respectively at low and high SNR regimes. As direct link is available, results indicate that $\mathcal{R}$ should be placed closer to $\mathcal{S}$. Though the user locations around $\mathcal{R}$ are more benefited, relay utility regions are not circular.}
 


\end{document}